\documentclass{nature}
\usepackage{epsf}
\usepackage{amssymb,amsmath}
\usepackage{multirow}
\usepackage{rotating}
\usepackage{subfigure}
\usepackage{subfig}
\usepackage{color}
\usepackage{deluxetable}
\usepackage{lipsum}

\newcommand\resetsubfigs{\setcounter{sub\@captype}{0}}

\def\aap{Astron.\ Astrophys.}
\def\apj{Astrophys.\ J.}

\def\apjl{Astrophys.\ J.}
\def\apjs{Astrophys.\ J.\ Supp.}
\def\mnras{Mon.\ Not.\ R.\ Astron.\ Soc.}
\def\araa{Annu.\  Rev.\ Astron.\ Astrophys.}
\def\aj{Astron.\ J.}

\def\pasj{Pub.\ Astron.\ Soc.\ Jap.}

\def\nat{Nature}
\def\nar{New\ Astron.\ Rev.}

\title{Discrete Knot Ejection from the Jet in a Nearby Low Luminosity Active Galactic Nucleus, M81*}

\author{Ashley~L.~King$^{1,2}$, Jon~M.~Miller$^{3}$, Michael Bietenholz$^{4,5}$, Kayhan G\"ultekin$^{3}$, Mark T. Reynolds$^{3}$, Amy Mioduszewski$^{6}$, Michael Rupen$^{7}$, Norbert Bartel$^{4}$ }

\begin{document}

\maketitle

\begin{affiliations}
\item { Department of Physics, 382 Via Pueblo Mall, Stanford, CA 94305, ashking@stanford.edu}
\item Einstein Fellow
\item Department of Astronomy, University of Michigan, 1085 S. University Ave, Ann Arbor, MI 48109-1107, USA
\item Department of Physics and Astronomy, York University, Toronto, M3J~1P3, Ontario, Canada
\item Hartebeesthoek Radio Observatory, PO Box 443, Krugersdrop, 1740, South Africa
\item National Radio Astronomical Observatory, P.O. Box O, Socorro, NM, 87801, USA 
\item{ NRC Dominion Radio Astrophysical Observatory, Penticton, British Columbia V2A 6J9}

\end{affiliations}

\begin{abstract}
Observational constraints of relativistic jets from black holes has largely come from the most powerful and extended jets\cite{Jorstad05,Asada14}, leaving the nature of the low luminosity jets a mystery\cite{Falcke04}. M81* is one of the nearest low-luminosity jets, which underwent an extremely large radio flare in 2011, allowing us to study compact core emission with unprecedented sensitivity and linear resolution. Utilizing a multi-wavelength campaign, we were able to track the flare as it re-brightened and became optically thick. Simultaneous X-ray observations indicated the radio re-brightening was preceded by a low energy X-ray flare at least $t_{\rm delay}>12\ {\rm days}$ prior. Associating the time delay between the two bands as the cooling time in a synchrotron flare\cite{Urry97,Bai03}, we find the magnetic field strength was $1.9<B<9.2\ {\rm G}$, which is consistent with magnetic field estimate from spectral-energy distribution modeling\cite{Kellerman81}, $B<10.2\ {\rm G}$. In addition, VLBA observations at 23 GHz clearly illustrate a discrete knot moving mildly relativistically at $v_{\rm app}/c=0.51\pm0.17$  associated with the initial radio flare. The observations indicate radial jet motions for the first time in M81*. This has profound implications for jet production, as it means radial motion can be observed in even the lowest-luminosity AGN, but at slower velocities and smaller radial extents ($\approx10^4\ R_{\rm G}$).
\end{abstract}

M81*, at a distance of $3.96\pm0.29$ Mpc\cite{Bartel07}, is one of the nearest low-luminosity active galactic nuclei (AGN), and has a black hole mass of $7^{+2}_{-1}\times10^7\ M_\odot$\cite{Devereux03}.  It has been well surveyed at many frequencies, including radio\cite{deBruyn76,Bartel82,Bietenholz04,Marti-Vidal11} and X-ray\cite{Ishisaki96,Merloni03,Markoff08}. Like many other low-luminosity AGN, the X-ray and radio luminosities of M81*, along with its mass, place it on the ``fundamental plane of black hole activity" \cite{Merloni03, Falcke04}. The existence of the this plane suggest a functional relationship between the jet production level, as indicated by the 5 GHz radio luminosity, the 2--10 keV X-ray emission, and the black hole mass\cite{Merloni03, Falcke04}. 

Although M81* fits on the fundamental plane within small fluctuations around its mean\cite{Miller10}, it shows highly variable radio emission\cite{Ho99} at frequencies greater than 5 GHz. At 15 GHz, M81* has an average flux density of 116 mJy\cite{Pooley11} with typical flares reaching on order of 150 mJy \cite{Ho99}.  Late in 2011, M81* exceeded its typical flare strength, increasing from a flux density of 140 mJy on 24 August 2011 (MJD 55797), to 261 mJy and 321 mJy on 27 August (MJD 55800) and 1 September (MJD 55805), respectively\cite{Pooley11}. Observers assembled a campaign in both the radio and X-ray bands, to track this substantial brightening, never before observed in a low-luminosity active galactic nucleus of such proximity. 

We targeted M81* with four epochs of radio observations, including both broadband frequency coverage with the {\it Karl Jansky Very Large Array} (VLA) and high resolution radio imaging with the {\it Very Long Baseline Array} (VLBA). Figures 1a \& 1b illustrate the VLA broadband radio frequency behavior in the unresolved compact core. As indicated by the decrease in flux density between September 17 and September 21, the flare was already cooling nearly 20 days after the initial report of activity. The radio spectral index, $\alpha$ (where $S_\nu\propto\nu^\alpha$), was negative, consistent with optically thin synchrotron emission. However, by the end of the four epochs, the radio emission at higher frequencies had re-brightened and the spectral index has become consistent with optically thick synchrotron emission ($\alpha >0$). This indicates renewed flaring activity at very small radii, i.e., the core of the jet. Very Long Baseline Array (VLBA) observations confirm this, and will be discussed below.

Compact radio core emission is generally modeled as self-absorbed synchrotron emission\cite{Cotton80}. The peak in the spectrum occurs at the boundary between optically thick and optically thin emission, from which we can make an estimate of the magnetic field strength\cite{Kellerman81}. Our first epoch observations peak at $\nu_{peak,1}=10.4^{+1.0}_{-2.5}\ {\rm GHz}$ with a flux density of $S_p =0.22\pm0.01\ {\rm Jy}$. Taking the restoring beam from the 8.4 GHz VLBA observations as an upper limit to the size of the core, we find the magnetic field strength to be $B\lesssim10.2\ {\rm G}$. See Supplementary Information section for more details. During the second epoch, the second flare peaks at roughly $\nu_p\sim28\ {\rm GHz}$, however with only a few data points at higher frequencies, it is difficult to assess the degeneracy between the peak of the second flare, the flux density, and the contribution from the emission at lower frequencies. However, higher frequencies reveal smaller spatial scales, closer to the base of the jet, indicating the second flare originated at smaller radii. In addition, changes in the spectral slope moving from optically thin ($\alpha<0$) to optically thick ($\alpha>0$) emission  at higher frequencies throughout the four epochs indicates increases in either the particle energy density or magnetic field strength (see Figure 1a \& 1b).

Contemporaneous X-ray observations also showed increased emission, which we associate with the radio re-brightening. In general, X-ray emission in low-luminosity AGN is modeled with a fiducial power-law attributed to either inverse Compton scattering of disk photons\cite{Merloni02}, synchrotron\cite{Markoff05} or synchrotron self-Compton\cite{Markoff08} emission from a jet. Shown in Figures  2a \& 2b are the 2011 absorption-corrected light curves in the 0.5-2 and 2--10  keV X-ray bands. The low energy band shows strong evidence for a flare at MJD=55816--55838, while the high energy band does not show as obvious of a trend. Due to lack of sampling prior to the start of the initial radio flare, it is unclear if a similar event preceded the initial radio flare. However, analysis of all the archival M81* {\it Swift} data finds this event to be unique in the six years of archival data (2006--2011), indicating its rare nature. See Supplementary Materials. As M81* is known to have many radio flares, though less intense, the rare X-ray behavior associated with this flare may explain its peculiar behavior, including the first detection of a discrete knot in M81*, which we discus in more detail below. 

Interestingly, the lack of a strong flare in the high energy band suggests that the ``corona", which is responsible for the production of the high-energy X-rays  via inverse Comptonization in AGN\cite{Merloni02}, is not associated with this particular radio flare. In addition, because the X-ray luminosity at 2--10 keV did not vary as dramatically as the emission at low frequencies, this suggest that even during this type of event, M81* would still lie on the ``fundamental plane of black hole activity''. Conversely, the low energy X-ray emission does appear to be associated with the radio flare at higher frequencies. It precedes the radio re-brightening by at least $t_{\rm delay}>12\pm1$ days, determined via a Z-transformed cross-correlation analysis. We note that the core is not in a flaring mode nearly three months later on January 16, 2012 based on its flux density and optically thin spectral index, which also sets an upper limit on $t_{\rm delay}$ of three months. This last epoch will be discussed below.

Accelerated charged particles in the jet core will cool via synchrotron radiation, and therefore the spectra will peak at progressively lower frequencies at later times. The time lag between the peak emission at different frequencies is dependent on the magnetic field strength, velocity of the flow, and viewing angle\cite{Urry97,Bai03}, as long as adiabatic losses do not dominate\cite{Jorstad05}. As will be discussed below, we find a one-sided VLBA radio knot moving with an apparent velocity of $v_{\rm app}/c=0.51\pm0.17$  and viewing angle to our line-of-sight of $\theta<56^\circ$. Assuming the de-projected velocity is close to the bulk flow velocity, we constrain the magnetic field strength, $B$, from the time lag between the X-ray flare and radio re-brightening to be $1.6<B<9.2\ {\rm G}$. We assumed the angle is $\theta=14^\circ$\cite{Devereux03} for the the upper limit, while the minimum flux ratio and maximum time delay give the lower limit. This is consistent with the magnetic field strength derived from the peak of the radio spectral-energy distribution in the first epoch, $B<10.2\ {\rm G}$. See the Supplementary Materials for further modeling details. 

Our measurement of the magnetic field is nearly two orders of magnitude greater than what is measured from the radio core-shift of M81*, which find a value on order of $B\sim34\ {\rm mG}$ at 0.21 milli-arcsec (1300 $R_{\rm G}$) scales at 8.4 GHz\cite{Marti-Vidal11}. Even if we extrapolate to 23 GHz, assuming the magnetic field is inversely correlated with radius, we find that the magnetic field strength measured from the core shifts is $B\sim90\ {\rm mG}$ at $\sim$0.08 milli-arcsec ($\sim$500 $R_{\rm G}$) , which is still two orders of magnitude less than our measurements at much smaller radii. Though this discrepancy could be due to measuring different components of the magnetic field via the two different methods, we are more likely measuring an increase in magnetic field associated with the large flare event. The magnetic field scales as $B\propto \nu_p^5 S_p^{-2} \theta_A^4$, where $\nu_p$ is the peak frequency, $S_p$ is the peak flux density, and $\theta_A$ is the angular size\cite{Kellerman81}. Even with such a large increase in magnetic field strength, only a small increase in either the peak frequency or emitting region is required in order to explain the increase in flux density observed during the flares. Both of which are reasonable assumptions, as we note the peak frequency does in fact increase in the re-brightening of the flare, and the emitting region extends, as evidenced by the detection of the knot discussed below.

We note that the radiation energy density determined via the radio luminosity of the knot is much less than the magnetic energy density as determined via synchrotron losses.  If the particle energy density is roughly in equipartition with the magnetic field, our results imply that the internal energy of particles and field dominate over the radiative energy, similar to what is observed in more massive AGN\cite{Merloni07} and stellar-mass black hole Cygnus X-1\cite{Gallo05}. 

In addition to having a strong magnetic field, we detect a discrete knot moving mildly relativistically for the first time in M81*. Figure 3a shows the knot as it moves radially outward over the four VLBA epochs, taken at 23.7 GHz with a full-width half-maximum (FWHM) restoring beam of 1.2$\times$0.55 milli-arcsecs with a position angle of $-23$ degrees. Figure 3b shows the 8.4 GHz observations, which have a larger restoring beam of 3.1$\times$1.4 milli-arcsecs with a position angle of 21 degrees that makes detection of the knot difficult in all but the last epoch. Figure 3c shows the spectral index, $\alpha$, between 8.4 and 23.7 GHz. At a distance of 3.96$\pm0.29$ Mpc\cite{Bartel07}, 1 milli-arcsec corresponds to $4.0\times10^3$AU, or equivalently, $5.7\times10^3R_{\rm G}$, where $R_{\rm G}=GM_{\rm BH}/c^2$ and $M_{\rm BH}=7^{+2}_{-1}\times 10^7\ M_\odot$, which indicates the knot has a projected distance of $R_{projected}\sim7500R_{\rm G}$. 

The elongation of the nucleus at 8.4 GHz has been observed before\cite{Bartel82,Bietenholz96,Bietenholz00,Bietenholz04,Markoff08,Marti-Vidal11}. However, we are the first to measure the radial motion of the knot in M81*. Fitting the knot detected at 23.7 GHz with a Gaussian component and determining its position with respect to the brightest component, i.e., the core, we find that the knot is moving radially from the center of the radio core with a projected velocity of $v_{\rm app}/c=0.51\pm0.17$  (see Figure 3a\&4a). Assuming the jet is bipolar, the apparent velocity, together with the brightness ratio lower limit from the one sided detection, place a constraint on the viewing angle of $\theta<56^\circ$ to our line-of-sight. See Figure 5.

This limit is broadly consistent with the angle of $\theta=14^\circ\pm2$ determined via optical modeling of the accretion disk on parsec scales\cite{Devereux03}. However, high resolution radio studies have found evidence that the core emission in M81* bends in the plane of the sky on milli-arcsec scales\cite{Bietenholz04} as well as precesses on a timescale of years\cite{Bietenholz00,Bietenholz04,Marti-Vidal11}, which suggests the inclination may change even on small scales, and long time periods. Deeper observations during similar outburst events are needed to further constrain the inclination angle. 

As the knot moves radially outward, its brightness also changes. Figure 4b shows the variability observed in both the knot and core at 23.7 GHz, as well as the variability observed in the core at 8.4 GHz. The core emission becomes optically thick by the end of the four epochs, which can also be seen in the spectral index map in Figure 3c. Figure 3c suggests that the knot is optically thin throughout the four epochs. The knot does however continue to brighten until the third epoch, and then drops in flux density by the fourth epoch. See Figure 4b. One can use the variability of the knot as it moves out along the jet to estimate the Doppler factor\cite{Jorstad05}, $\delta_{\rm Doppler}$, shown as the red curve in Figure 5. We discuss this in more detail in the method section, and find the Doppler factor to be $\delta_{\rm Doppler}\sim1.5$. This corresponds to a viewing angle of $\theta\sim44^\circ$, which is much larger then the angle inferred from optical modeling. However, large uncertainties due to the poor temporal sampling of the knot light curve do not exclude a $\theta=14^\circ$.

We also reduced archival VLBA observations of M81*  on January 16, 2012 (MJD 55943), nearly three months later, which also show a discrete knot, in addition to the compact core emission at both 23 and 15 GHz  (Figure 6a--e). Again, due to the lack of temporal sampling, we can not distinguish between a separate, distinct knot ejection associated with the re-brightening of the core, stalling of the original knot, or a recollimation of the jet.  The January knot might have been launched during the radio re-brightening flare if the velocity is less than that observed from the first knot.  In contrast, the knot may have decelerated to the position in this fifth epoch. Recollimation of jets and deceleration of knots is observed in other jets, like M87\cite{Asada14}, but at much larger gravitational radii ($\sim 10^6 R_{\rm G}$) compared to what is observed in M81*. 

The closeness of the bright knots to the core in M81*, $R<10^5\ R_{\rm G}$ (for all but the smallest inclinations, $\theta\lesssim6^\circ$) is peculiar as compared to other AGN of similar masses but which are radio loud, e.g., 3C 120 and 3C 111\cite{Chatterjee09,Chatterjee11}. These latter AGN produce jets that  reach radial extents exceeding $R>10^6R_{\rm G}$. The mildly relativistic knot velocity may partially explain the small radial extent of M81*, as the knot may not travel as far before dissipating.  However, this can not be the only solution, as M87 has both mildly relativistic knots at small radii and highly relativistic knots at large radii\cite{Asada14}. Therefore, high magnetic field strength, accretion rate, small black hole mass, or even a potentially low spin may be responsible for setting the small radial extent in M81*.

Current paradigms suggest that jet production in both stellar-mass and supermassive black holes at low mass accretion rates should be steady and continuous\cite{Falcke04}, while black holes with high mass accretion rates, $L_{\rm X}\sim L_{\rm Edd}$ -- produce jets with discrete, highly relativistic knots\cite{Wehrle01,Chatterjee09,Chatterjee11}. As M81* has a relatively low-mass supermassive black hole ($7\times10^7M_\odot$\cite{Devereux03}) and is accreting material at a very low rate ($L_{\rm X}\sim10^{-5.8} L_{\rm Edd}$), detecting discrete, mildly relativistic knots reveals the similarities between jet production at {\it all} masses and mass accretion rates. The observations demonstrate that though the radial extent and knot velocities may differ, jets are intrinsically capable of producing the same knot-like structures. In addition, the multi-wavelength analysis indicates that jet knot ejections are not only associated with radio flares, but also X-ray flares, and can be further utilized to measure the magnetic field strength very close to the black hole.

{\bf Author Contributions:} A.L.K led the data reduction and analysis, with contributions from J.M.M., M.B., A.M..  K.G., M.T.R., M.R. and N.B contributed to discussion and interpretation.

{\bf Corresponding author:} Correspondence and requests for materials should be addressed to A.L.K. (ashking@stanford.edu).

{\bf Acknowledgements:} The authors would like to thank the referees for their invaluable comments. ALK would like to thank the support for this work, which was provided by NASA through Einstein Postdoctoral Fellowship grant number PF4-150125 awarded by the Chandra X-ray Center, which is operated by the Smithsonian Astrophysical Observatory for NASA under contract NAS8-03060
\newpage
\begin{figure}
\label{knot2}
\captionsetup{type=figure}
\includegraphics[width=.95\linewidth,angle=0,clip=true,trim= 0 -80 -2cm -5cm]{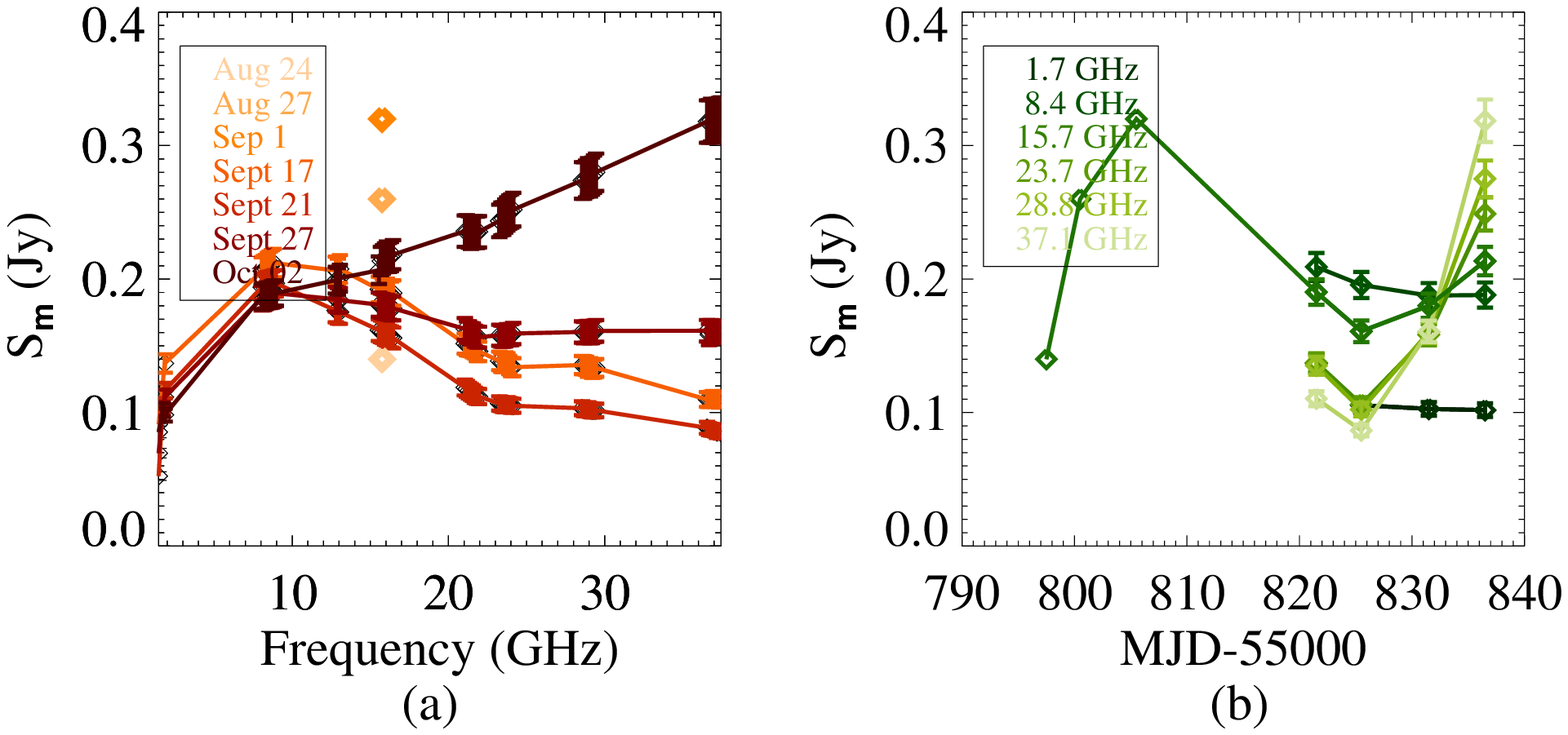}
\caption{ {\bf Radio Flare Spectral Energy Distribution:} Flux density measurements of M81* at eight radio frequency bands between 1.7 and 37.1 GHz. Our own observations were made with the VLA at four epochs from September 17, 2011 (MJD 55821.7) to October 2, 2011 (MJD 55836.7), and we also plot the earlier 15 GHz observations of Pooley\cite{Pooley11}, made with the Arcmintue Microkelvin Imager of the Mullard Radio Astronomy Observatory. Panel {\bf a} shows the radio spectral energy distribution at different epochs, while panel {\bf b} shows the flux density of different frequencies as a function of time. Between September 17 and September 27, the spectrum turns over at $\sim$10~GHz, suggesting that it is optically thin above that frequency, and that the flare is decaying. However, on October 2, the whole spectrum is inverted, suggesting that it has becoming optically thick and that the flare is rebrightening at least at the higher frequencies. Error bars are 1$\sigma$ confidence.}
\end{figure}
\newpage

\setcounter{subfigure}{0} 
\begin{figure}
\label{xray}
\captionsetup{type=figure}
\includegraphics[width=.95\linewidth,angle=0,clip=true,trim= 0 0 -1cm -1cm]{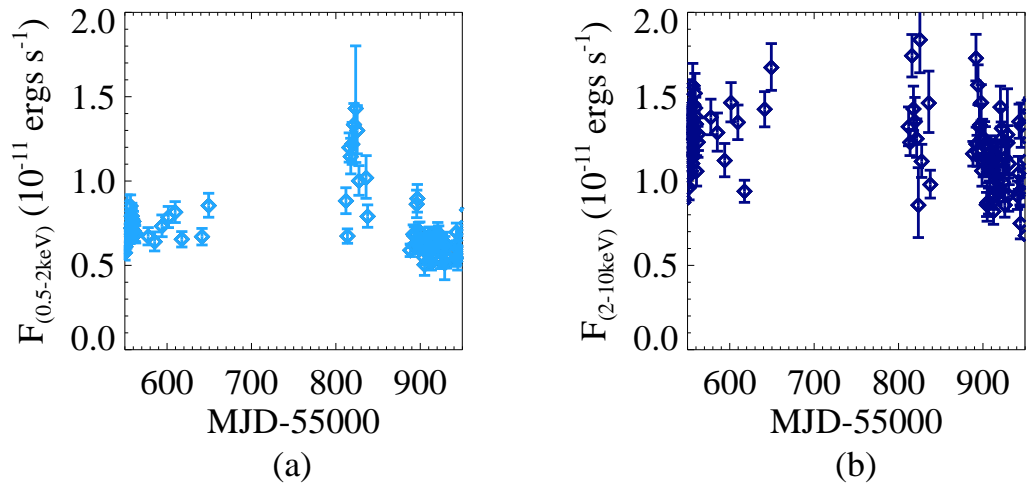}
\caption{{\bf X-ray Flare:} The X-ray flux as a function of time, observed with the {\it Swift} satellite. Panel {\bf a} shows the absorption-corrected flux in the low-energy, 0.5--2 keV X-ray band. The flux remains relatively constant except for the flare which occurs during MJD=55816--55838. Panel {\bf b} shows the absorption-corrected flux in the high-energy, 2--10 keV x-ray band. There is no obvious flare in this band that can be associated with either the flares in the radio (see Figure 1) or the low-energy  X-ray flare. Error bars are 1$\sigma$ confidence. }
\end{figure}

\setcounter{subfigure}{0} 
\begin{figure}
\captionsetup{type=figure}
\vspace{0cm}
\includegraphics[width=.95\linewidth,angle=0,clip=true,trim= -40 -60 -1cm -5cm]{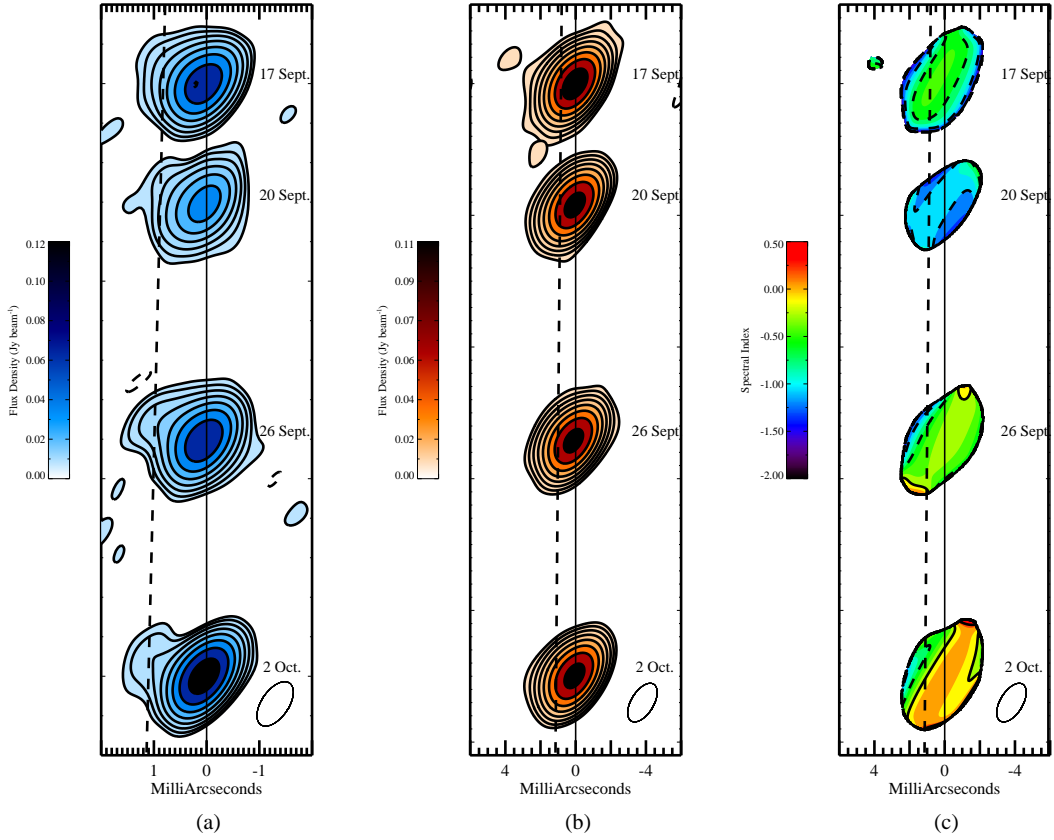}
\caption{ {\bf High Resolution Radio Knot Motion:} The three panels show the AGN core emission from M81* at 23.7, 8.4 GHz, and the resulting spectral index ($S_\nu \propto \nu^\alpha$) on 17 September (MJD 55821.7), 20 September (MJD 55824.8), 26 September (MJD 55830.8), and 2 October (MJD 55836.7) of 2011. The contour levels in panel a) is $3.4\times(-3, 3, 6,$ $12, 24, 48, 96, 192, 384)\times10^{-4} {\rm Jy~beam}^{-1}$ with a restoring beam size of $1.2\times0.55$ milli-arcsec with a position angle of $-23.0^\circ$. The contour levels in panel b) is $2.9\times(-3, 3, 6,$ $12, 24, 48, 96, 192, 384)\times10^{-4} {\rm Jy~beam}^{-1}$ with a restoring beam size of $3.1\times1.4$ milli-arcsec with a position angle of $-21.0^\circ$. And the contours in panel c) are $(-3,-2.5,-2,-1.5,-1,-0.1,0,0.5,1)$ with the negative values in dashed lines and the restoring beam of $3.1\times1.4$ with a position angle of $-21.0^\circ$. The solid lines denotes the position of the core emission. A knot is clearly detected in all four epochs of the 23 GHz. Gaussian fits suggest a velocity of $v_{\rm app}/c=0.51\pm0.17$ , given by the dashed line. In addition, the core emission moves from a negative index in the first three epochs, to a positive index in the last epoch as it becomes optically thick during the beginning of the second flare. There is a hint that the knot emission has a negative spectral index even during the last epoch, suggesting it is optically thin as it propagates outward. Error bars are 1$\sigma$ confidence. \label{fig:epochs}}
\end{figure}

\newpage

\setcounter{subfigure}{0} 
\begin{figure}
 \captionsetup{type=figure}
 \includegraphics[width=.95\linewidth,angle=0,clip=true,trim= 0 -10cm -3cm -10cm]{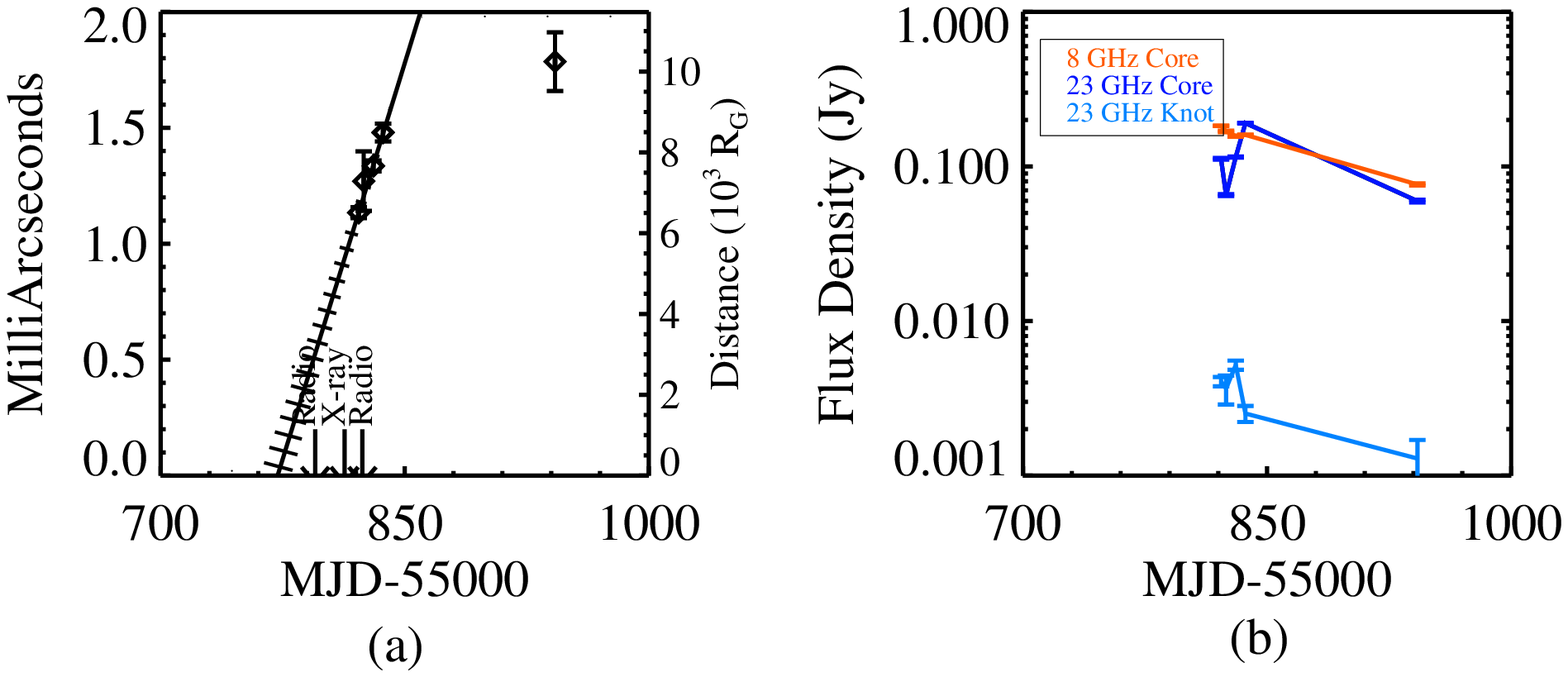}
\caption{{\bf Changes in Knot Position and Brightness: } Panel {\bf a} shows the relative position of the knot. The linear fit to the first four epochs (solid line) corresponds to an apparent velocity of $v_{\rm app}/c=0.51\pm0.17$. Due to the close proximity of M81*, we are able to resolve discrete knot ejections in a compact radio core at high radio frequencies. The fifth data point is possibly associated with the second radio flare or a recollimation shock.  Panel {\bf b} shows the 23.7 GHz flux density of the knot (light blue) as well as the flux density of the core at both 23.7 GHz (dark blue) and 8.4 GHz (red).Error bars are 1$\sigma$ confidence. \label{fig:vel} }
\end{figure}

\newpage

\begin{figure}
{\includegraphics[width=.6\linewidth,angle=0,clip=true,trim=2cm 0cm -5cm -10cm]{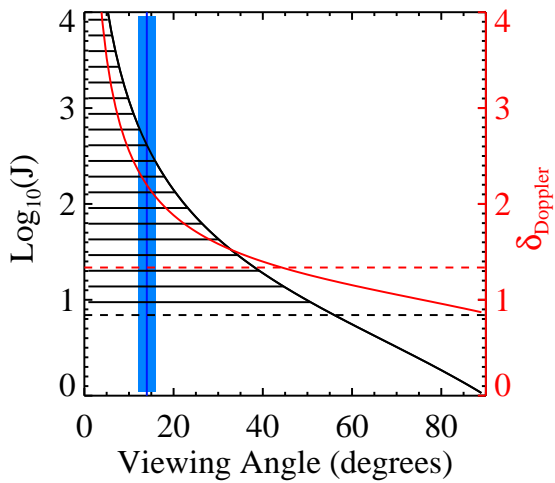}}
\caption{{\bf Constraints on Orientation:} This figure shows Brightness ratio (J) of the approaching and receding knots as a function of viewing angle in black. The dashed black line shows the lower limit to the ratio of the brightest knot on the third epoch, which puts a limit of the viewing angle of $\theta<56^\circ$. In red is the Doppler factor ($\delta_{\rm Doppler}$) as a function of viewing angle. We measure a Doppler factor from the knot variability during the four epochs shown as the red dashed line, which corresponds roughly to an inclination of $\theta\sim44^\circ$. The blue vertical line shows the inclination of $\theta=14^\circ\pm2$, measured from spectral fitting\cite{Devereux03}.  Errors on the inclination are 1$\sigma$. \label{fig:ang} }
\end{figure}
\vspace{2cm}
\pagebreak

\setcounter{subfigure}{0} 
\begin{figure}
\captionsetup{type=figure}
\includegraphics[width=.95\linewidth,angle=0,clip=true,trim= -10 -10 -1cm -5cm]{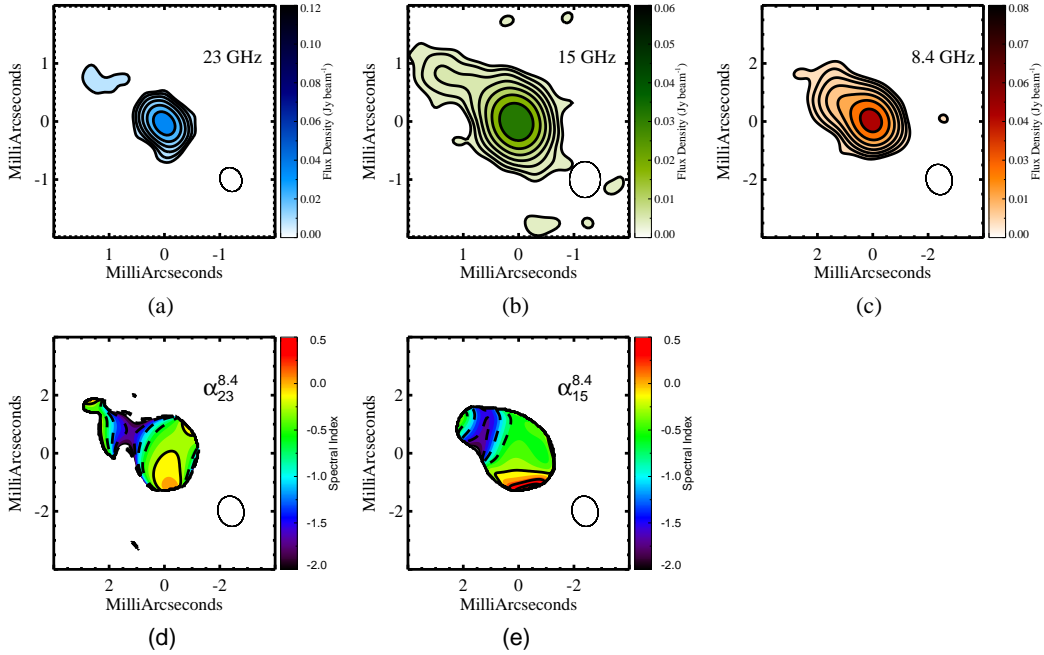}

\caption{{\bf VLBA 16 January 2012  Observations:} (MJD 55943) This figure shows the 23.7 GHz, 15 GHz, 8.4 GHz, and spectral index maps for the fifth epoch. Panels have contours of the $\sigma_{rms}\times(-3, 3, 6,$ $12, 24, 48, 96, 192, 384)\times10^{-3} {\rm Jy~beam}^{-1}$, where $\sigma_{rms}$ is 0.32, 0.15 and 0.22 mJy beam$^{-1}$, for a, b and c respectively. Panels c and d have contours of $(-3,-2.5,-2,-1.5,-1,-0.1,0,0.5,1)$.  The restoring beams are shown in the bottom right corner; a) $0.43\times0.38$ milli-arcsec with a position angle of 36$^\circ$, b) $0.62\times0.56$ milli-arcsec with a position angle of $-0.6^\circ$, and c-d) $1.1\times0.93$ milli-arcsec with a position angle of 21$^\circ$. As can be seen in a-c, the knot is extended much further from the core than in previous observations. Panels d and e, show the core and knot are both optically thin ($\alpha<0$) indicating the flare has stopped.}
\end{figure}


{\large \bf Supplementary Information}

{\bf Radio Reduction}

M81* exhibited a large radio flare on 23 August 2011\cite{Pooley11}, reaching 321 mJy by 1 September 2011\cite{Pooley11}. The mean flux density during the previous 5 years was 116 mJy, at 15 GHz. We triggered four epochs of observations with the VLA on 17 September, 21 September, 26 September, and 2 October 2011. During the first two observations, the telescope array was in transition from its most extended configuration (A) to its most compact configuration (D). The later two observations were both taken in the compact, D configuration. We used radio bands centered on the frequencies of 1.2, 1.7, 8.5, 11.9, 12.6, 16.0, 21.5, 23.7, 29.0, and 37.0 GHz. Each sub-band had a bandwidth of 1.024 GHz, except the 1.7 GHz which had a 512 MHz bandwidth. 

3C 286 was used as the flux density calibrator with two minute integrations at each band, except 1.7 GHz, which had 3.5 minutes. J0958+6533 was used as the primary phase calibrator, bracketing the integrations on M81*. A total of seven minutes was spent on M81* at each frequency band, except at 1.7 GHz, which had 5.25 minutes on source. The pointing offsets were calibrated before the high frequency observations. The total observing time at each epoch was 1.5 hours.

Utilizing the CASA software package, version 3.4.0\cite{McMullin07}, the visibilities were examined for radio frequency interference, which primarily affected the low frequency bands. Standard reduction techniques were used to transfer the flux density scale from the flux density calibrator to phase calibrator and M81*, and then bandpass and gain calibrations were done. A Gaussian centered at the phase center was then fit to the uv visibilities. A 5\% error was added to the statistical fit error to account for the uncertainty in the absolute flux density scale. Figures 1a \& 1b show the resulting flux density measurements. 

We also triggered observations at 8.4 and 23 GHz with the VLBA. These observations were taken on the same days as the VLA observations, except for the second epoch which was taken on 20 September 2011. We focus the following method sections on the 23 GHz data reduction, though the reduction was similar for 8.4 GHz data. In each observation epoch, 90 minutes were spent on source, M81*, and 50 minutes on the phase calibrator, J0958$+$6533. The observations were centered at 23.7 GHz, and had 8 spectral windows, each with 16 spectral channels. Each spectral window had 8 MHz of bandwidth giving a total bandwidth of 64 MHz, and were taken with full polarizations.  Each epoch had 8 telescopes in the array, except on September  20, which had 6.  

We reduced the data using NRAO's {\it AIPS} software, version 31DEC13. We again used J0958+6533 as the phase calibrator. A grid of 512$\times$512 pixels with size of $4.5\times10^{-2}$ milli-arcsec was used to produce an image. After the initial gain calibrations were transferred to M81*, we proceeded to use self-calibration to improve the images.  Initial self-calibration was phase-only. Amplitude and phase calibrations were then carried out with solution intervals initially set to the entire observation, decreasing with each iteration. Figure 3a shows the 23.7 GHz VLBI image of M81*. The resulting root-mean-square (rms) of the images is given in Table 1. Figures 3b and 3c show the 8.4 GHz VLBI images and spectral index map of M81*, respectively.

The emission at 23.7 GHz was modeled in the UV plane with {\it OMFIT} using four point sources (see Figure 4a, Table 1). The first two adjacent point sources were used to fit the core emission, and two additional point sources were used to model the knot as well as the extended emission observed to the north-east in each image. Though {\it OMFIT} produces an error associated with the measured positions, we calculated the position errors using the full-width at half-maximum of the beam divided by the component's signal-to-noise, which increased the errors, allowing a more conservative estimate fore the measured velocity. The third component, between the core and the most extended emission, is the least constrained, likely due to the approximation of the irregular structure with Gaussian components. After the positions had been measured, distances between the brightest knot and the brightest core component were computed. The westernmost core component was always the brightest one. We note that in the 8.4 GHz observations, the extended emission did not extend farther than 2 milli-arcsec ($1.1\times10^4R_{\rm G}$). See Figure 3b.

Utilizing the projected distance measurements for the discrete knot moving to the north-east, we made a linear fit to the distance as a function of time. We found that the apparent velocity was $v_{\rm app}/c=0.52\pm0.08$ ($v_{\rm app}=8.2\pm1.0$ milli-arcsec year$^{-1}$, $1\sigma$ confidence errors). When calculating the uncertainty in the apparent velocity in physical units, we include both the statistical errors from the linear fit to the distance in Figure 4a, as well as the distance uncertainty to M81* of 3.96$\pm$0.29 Mpc\cite{Bartel07}. There may be an intrinsic bias in the first observations when the knot is initially detected. The knot emission may still overlap with the core emission, which would bias the measured position toward the core. If only the first or second observations were effected, i.e., the observations where the knot is closest to the core, then this would artificially increase the measured velocity ,as it would  artificially decrease the distance measured from the core in these epochs. We therefore exclude the first epoch from our velocity fits, measuring a velocity of  $v_{\rm app}/c=0.51\pm0.17$. More conservatively, if we exclude both the first and second epochs, we measure a velocity of $v_{\rm app}/c=0.55\pm0.18$.  These velocities are still consistent with our initial measurement and $3\sigma$ inconsistent with a zero velocity, indicating the knot is not stationary, and moving radially from the core. Consequently, we adopt a value of $v_{\rm app}/c=0.51\pm0.17$ as the measured velocity, as it is consistent with all the velocity fits, but likely best estimates the systematic uncertainties of the knot velocity measurement.

Finally, the spectral index maps displayed in Figures 3c, 6d and 6e were made by convolving the highest resolution restoring beams with the lowest resolution restoring beams (8GHz). Then we utilized the {\it AIPS} tool {\it COMB} to find the spectral index between the two frequency across the image, making sure to clip the initial images below three times the noise level to ensure a significant detection of the spectral slope. Some emission at higher frequencies could be resolved-out when comparing to the 8.4 GHz observation, making the spectral index a lower limit. However, we have quantified the amount of flux density that could be resolved-out using the tool {\it UVSUB} to transfer the 8 GHz cleaned model components through the 23 GHz UV coverage. After convolving the transformed 8 GHz visibilities with the initial 8.4 GHz restoring beam, we compared the resulting image to the original 8.4 GHz image. We find that only 1.9\% of the flux density is resolved-out during this last epoch.

{\bf X-ray Data Reduction}
The X-ray data were taken with the {\it Swift} satellite between September 7, 2011 to October 3, 2011, with a cadence of approximately two days. Additional {\it Swift} data from 2011 were also included for comparison to the campaign during the radio flare. The average count rate for all the observations in 2011 was 0.42 counts s$^{-1}$ with an average exposure time of 960 seconds. 

The data were reduced using the {\it xrtpipeline} via {\it FTOOLS} software, version 6.16\cite{Blackburn95}. Exposure maps and background files were created using the standard tools {\it xrtexpomap} and {\it xselect}, respectively. Extraction regions for the source were nominally an annulus with an outer radius of 118 arcsecs and an inner radius of 4.7 arcsec to avoid pile-up contamination. The background region also used a annulus with outer radius of 213 arcsec and inner radius of 141 arcsec. 

After the source and background spectra were extracted, we utilized {\it XSPEC}, version 12.8.2\cite{Arnaud96}, to fit the data with a neutral absorption component modeled with {\it tbabs}, and a power-law component. The absorption component was frozen at the Galactic hydrogen column density of $5\times10^{20}$ cm$^{-2}$  \cite{Kalberla05}, while the power-law index and normalization were allowed to vary.  Figures 2a \& 2b show the low-energy flux (0.5--2.0 keV), and high-energy flux (2-10 keV) during 2011. The high energy mean for 2011 was $1.2\times10^{-11}$ ergs s$^{-1}$ cm$^{-2}$ with a standard deviation of $0.2\times10^{-11}$ ergs s$^{-1}$ cm$^{-2}$. In the low energy band (0.5--2 keV), excluding the time of the flare ($55790-55840$ MJD), the data fluctuate around a mean of $6.3\times10^{-12}$ ergs s$^{-1}$ cm$^{-2}$ with a standard deviation of $0.8\times10^{-12}$ ergs s$^{-1}$ cm$^{-2}$. The low-energy X-ray flaring data had a peak well above the mean at $1.4^{+0.4}_{-0.3}\times10^{-11}$ ergs s$^{-1}$ cm$^{-2}$. 

We show that this flare is model independent by plotting the intensity across the entire band (0.5--10 keV) versus the ratio between the detected counts in the 0.5--2 keV and 2--10 keV bands in Supplementary Information Figure 1a. The flare, depicted in black, moves well above all the archival {\it Swift} data to lower energies as the flare proceeds (clockwise). Though the flare is not strictly unique in this behavior compared to other outliers in Supplementary Information Figure 1a, it does show unique behavior in terms of the overall intensity. We preform a 2D Anderson-Darling test utilizing the python {\it scipy} (version 0.14.0) script {\it anderson\_ksamp}  to determine how unique the flare intensity distribution is compared to the rest of 2011 and to the entire six year period. We find that the flare distribution is inconsistent with both the 2011 and the six year intensity distribution at the $4.3\sigma$ and $4.0\sigma$ level.

Moreover, we fit the data with a simple power-law model shown in Supplementary Information Figure 1b. The average spectral index is $\Gamma=1.72\pm0.14$, while the peak spectral index is $\Gamma=2.41\pm0.42$. We find this event to be rare, in that analysis of all the M81* {\it Swift} archival data (2005-2011) shows only this one event is associated with a statistically significant ($3\sigma$) increase in the spectral index above the nominal $\Gamma=1.7$. Interestingly, an increase in the spectral index to $\Gamma\simeq2.5$ is also seen in the black hole at the center of the Milky Way, Sgr A*, during its X-ray flares\cite{Porquet03}. An increase in spectral index and X-ray brightening also occurs in stellar-mass black holes before discrete knot ejections as well\cite{Fender04}. This suggest that this behavior is not unique to supermassive black holes but is ubiquitous in jet production in accreting black holes across the mass scale. 

If we add a second neutral absorption component that is allowed to vary to account for intrinsic variable absorption in M81*, the results do not vary significantly. A varying absorption component could easily explain the other outliers in Supplementary Information Figure 1a as absorption moves out of our line-of-sight, increasing the counts in the soft band, though not necessarily the spectral index. This indicates that the low-energy X-ray flare is intrinsic to the source and not to neutral material serendipitously moving out of our line-of-sight. 

\setcounter{subfigure}{0} 
\begin{figure}
 \captionsetup{type=figure}
 \includegraphics[width=.95\linewidth,angle=0,clip=true,trim= 0 0 -1cm -5cm]{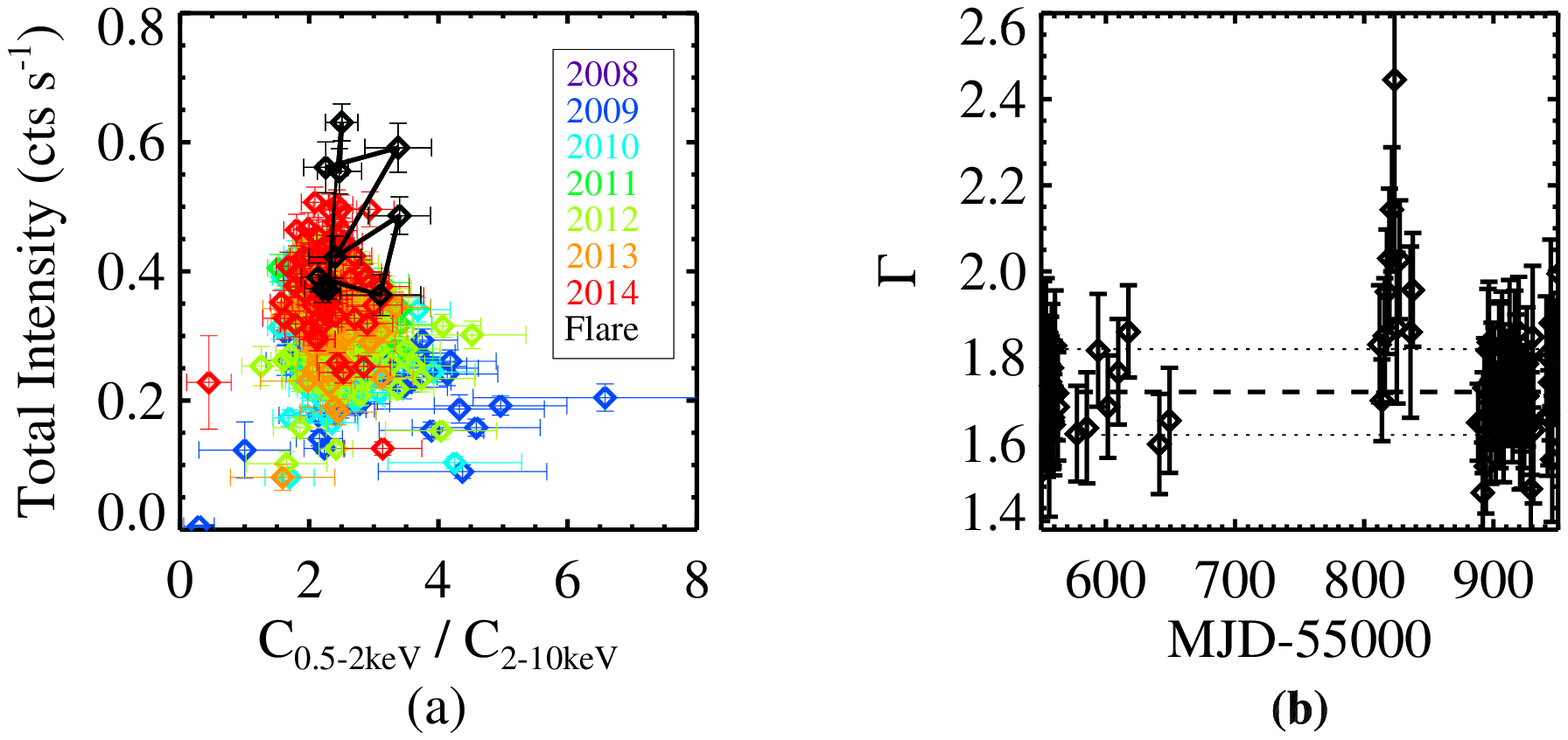}
\caption{{\bf X-ray Flare Characteristics:} Panel a: This figure shows the count rate versus the count rate ratio between the 0.5--2 keV and 2--10 keV X-ray bands. This shows that the flare (in black) is model independent, increases in count rate well above the typical count rate, while also becoming ``softer" dominated by the low energy band emission. The flare moves clockwise in this diagram. Panel b: The X-ray spectral index, measured in the 0.5-10 keV band, increases during the soft-energy X-ray flare, and then returns to its nominal value. The dashed line indicates the average spectral index, excluding the flare from the fit. The dotted lines are 1$\sigma$ standard deviations. There is no other event that deviates from the average spectral index by over $3\sigma$ confidence in all of the {\it Swift} archival data. Error bars are 1$\sigma$ confidence. }
\end{figure}

{\bf Cross Correlation Analysis}
In order to estimate the time delay between the low-energy X-ray and radio flare, we utilized a Z-transformed discrete correlation function\cite{Alexander13}. This analysis allowed for the cross-correlation between two light curves with sparse, uneven temporal sampling. We required that at least three points be used in the the time lag bin. Cross-correlation between the low-energy X-ray band (0.5-2.0 keV) and the three radio bands, 15, 23 and 37 GHz indicate a peak 12$\pm$1 days in all three bands as we do not yet measure the peak of the flare in any of these three bands. A positive lag means that the X-rays are leading the radio flare. The uncertainty on this time delay is strictly statistical and does not take into account systematics effects that can result from under-sampling the light curves. Likewise, as we do not measure a peak in the flux density, our measurement is a lower limit, $t_{\rm delay}>12\pm1$ days. However, we can place a large upper limit to the time delay, as we know the flare to have finished by the fifth epoch of observation with the VLBA on 16 January, 2012. Therefore the time delay is between $12<t_{\rm delay}<106$ days.

{\bf Magnetic field Estimates}
The magnetic field strength can be estimated using the time delay between the X-ray and Radio bands\cite{Bai03,Urry97}, with the following equation.

\begin{equation}
t_{\rm delay}\approx2\times10^7 [(1+z)/\delta_{\rm Doppler}]^{-1/2} B^{-3/2} [{\rm s}]
\end{equation}

where $t_{\rm delay}$ is the time delay between X-ray and radio bands, $\delta_{\rm Doppler}= [\gamma(1-\beta\cos\theta)]^{-1}$is the Doppler factor, $\gamma$ is the Lorentz factor, $\beta$ is the intrinsic velocity, $\theta$ is the viewing angle, and $B$ is the magnetic field strength measured in Gauss. This method assumes that both the X-ray and radio flares are produced in the same shock event, and both frequency bands are dominated by synchrotron emission. However, if the X-rays are instead dominated by synchrotron self-Compton scattering\cite{Markoff08}, we would expect the X-rays to lag behind the radio flare\cite{Marscher01}.

We estimate the Doppler factor with three methods, all of which utilize measurements of the knot observed with the VLBA. We caution that this knot was ejected during a previous flare, but is assumed to be representative of subsequent flares. The first method of estimating the Doppler factor assumes the jet is bipolar, and uses the ratio of flux densities between the approaching and receding knots via the following relation,
\begin{equation}
J=\left(\frac{1+\beta\cos\theta}{1-\beta\cos\theta}\right)^p
\end{equation}
where $J$ is the brightness ratio, $\beta$ is the actual velocity $\beta=\beta_{\rm app}/(\beta_{\rm app}\cos\theta+\sin\theta)$, $\theta$ is the angle to our line-of-sight, and $p=3-\alpha$, where $\alpha$ is the radio spectral index. We assumed the knot emission was optically thin and discrete, i.e., $\alpha\approx-0.7$. See spectral index maps Figures 3c. Due to our lower limit on the brightness ratio, we place an upper limit on the inclination of $\theta<56^\circ$, and a lower limit on the Doppler factor of $\delta_{\rm Doppler}>1.24$. See the black curve in Figure 5. This also assumes that once the apparent velocity of the knot is de-projected to intrinsic velocity of the knot, it is the bulk flow speed.

The second method for estimating the Doppler factor utilizes the flux density variability observed in the knot, and assumes such variability occurs on the light travel time across the knot \cite{Jorstad05}. 
\begin{equation}
\delta_{\rm var} = \frac{s {\rm D}}{c \Delta t_{\rm var} (1+z)}
\end{equation}
where $\delta_{\rm var}$ is the Doppler factor derived from variability, $s$ is the angular size of the component and we assumed 1.6$a$, where $a$ is the full-width half-maximum of the Gaussian components. Further, D is the distance to M81*, $c$ is the speed of light, and $\Delta t_{\rm var}$ is the variability timescale defined as $\Delta t_{\rm var}=dt/\ln(S_{\rm max}/S_{\rm min})$. Here, $dt$ is the time between the maximum ($S_{\rm max}$) and minimum ($S_{\rm min}$) flux densities. This methods assumes radiative losses, and we note that the knot we observe may be subject to adiabatic losses in addition to radiative losses. However, variability studies of quasars show that radiative losses are the dominant cooling mechanism in similar knot structures\cite{Jorstad05}.

 As shown in Figure 4b, the knot peaks in flux density during the third epoch on 26 September 2011. Almost a week later it shows a minimum in flux density on 2 October 2011. This gives a brightness ratio of $S_{\rm max}/S_{\rm min} = 2.1\pm0.6$ and a time delay of $dt = 6$ days. The brightest knot was fit with a Gaussian with full-width half-maximum of $0.76\times0.29$ milli-arcsecs with a position angle of $-16.4$ degrees. The shortest light travel path is across the minor axis, thus we used $a=0.29$ milli-arcsec for the angular size of the knot, which gives an estimate of $\delta_{\rm var} = 1.33$. If we assume that the Doppler boosting observed from the knot is equal to the Doppler boosting set by the variance, this corresponds to a viewing angle of $\theta = 44^\circ$, which is broadly consistent with the inclination inferred from the brightness ratio of approaching and receding knots. See the red curve in Figure 5.
 
Unfortunately, the temporal sampling is sparse, resulting in a large uncertainty in both the time as well as the ratio between the maximum and minimum flux densities. By definition, the observed ratio between the maximum and minimum is a lower limit, making $\delta_{\rm var}$ a lower limit as well. However, the time between the maximum and minimum could both be shorter or longer, resulting in a larger or smaller $\delta_{\rm var}$, respectively. Therefore, estimating the uncertainty in our measurement of $\delta_{\rm var}$ is difficult and should be taken as an initial estimate only.

Finally, we utilize the viewing angle of $\theta=14^\circ\pm2$, inferred from optical modeling of the M81* accretion disk \cite{Devereux03}, as well as the apparent knot velocity of $v_{\rm app}/c=0.51\pm0.17$  as a third method to infer a Doppler factor of $\delta_{\rm Doppler}=2.2\pm0.4$. See blue curve in Figure 5. This is also broadly consistent with the estimates inferred from our previous two methods. However, we stress that the viewing angle is highly model dependent, and although it gives low statistical errors, this inclination measurement may also be subject to systematic errors that could bias it both higher or lower.

Moving forward, we assume the Doppler factor from the second flare is the same as the first, which is derived from the observed knot and its structure. We use $\delta_{\rm Doppler}=2.2$ in the magnetic field calculation, as it is our largest estimated Doppler factor, while still consistent with all three methods. As the time-delay is assumed to be a lower limit, due to the failure to sample the peak of the radio flare, both quantities, $\delta_{\rm Doppler}$ and $t_{\rm delay}$, and their limits make the estimate of the magnetic field an upper limit (i.e., $B\propto \delta_{\rm Doppler}^{1/3} t_{\rm delay}^{-2/3}$). We estimate the magnetic field strength to be $B<9.2\ {\rm G}$. If we assume that the time delay is an order of magnitude larger, $t_{\rm Delay}<106\ days$, which is roughly the time of our fifth epoch, where we do not see the core in a flaring state, and the Doppler factor has its minimum possible value of $\delta_{\rm Doppler}>1.4$, we find the minimum magnetic field strength of $B>1.9\ {\rm G}$.

The magnetic field strength can also be assessed utilizing both the angular size and spectral-energy distribution of the core emission. This is done via the following relation\cite{Kellerman81},
\begin{equation}
B \sim f^{-5}\nu_p^{5}S_p^{-2}\theta_A^{4}(1+z)^{-1} [G]
\end{equation}
 where $f\sim 8$ for an electron distribution with an index of $\gamma=2$, ($N(E)dE\propto E^{-\gamma}dE$), $S_p$ is the flux density in Jy at the peak frequency, $\nu_p$, measured in GHz. To find the peak of the spectral energy distribution during our first epoch of observation we extrapolate the spectrum from both low and high frequencies. We fit the spectrum between 1.8 and 8.4 GHz with a exponential fit of $S_\nu \propto \nu^\alpha$, where $\alpha=0.27\pm0.04$. This is consistent with the slope measured just with the 8.4 GHz. At higher frequencies, we find the best fit between 12 and 16 GHz with exponential fit has an exponent of $\alpha=-(0.35\pm0.27)$. These fits intersect at $10.4^{+1.0}_{-2.5}$  with a peak flux density of 0.22$\pm0.01$ Jy. Assuming the angular size at 10.4 GHz is smaller than the restoring beam which scales inversely with frequency, our smallest full-width half-maximum restoring beam at 8.4 GHz of 0.75 milli-arcsecs puts an upper limit of $B\lesssim10.2$ G on the magnetic field strength. This is of the same order of magnitude as our previous estimate using the time delay and Doppler boosting of the knot.

\begin{deluxetable}{c c | c c c c c | c c c c c } 
\tabletypesize{\scriptsize}
\tablecolumns{7} 
\tablewidth{0pc} 
\renewcommand{\arraystretch}{0.6}
\tablecaption{ VLBA Component Parameters} 
\tablehead{ & & \multicolumn{5}{c}{23 GHz} &  \multicolumn{5}{c}{ 8.4 GHz}  \\
\colhead{Date} & \colhead{Comp} & \colhead{S$_\nu$} & \colhead{D} & \colhead{$\sigma_{\rm rms}$} & \colhead{$\theta_{\rm FWHM}$}& \colhead{$\theta_{\rm PA}$}  & \colhead{S$_\nu$} & \colhead{D} & \colhead{$\sigma_{\rm rms}$} & \colhead{$\theta_{\rm FWHM}$}& \colhead{$\theta_{\rm PA}$}\\
& & (mJy) & (marcsec) & (mJy & (marcsec) & ($^\circ$) & (mJy) & (marcsec) & (mJy & (marcsec) & ($^\circ$) \\
& &  & & beam$^{-1}$) & & & & & beam$^{-1}$) &   }
\startdata
Sept. 17 & & & &0.21 & 0.78$\times$0.32 &  5.7$^\circ$ & & & 0.13 &  2.3$\times$0.72 & 0.8$^\circ$\\
&       1 & $  62.4\pm  1.1$ & - & &  & &  $ 103.0\pm  0.5$ & - \\
&       2 & $  49.6\pm  1.0$ & $  0.18\pm  0.01$ & &  & &  $  80.2\pm  0.5$ & $  0.43\pm  0.01$ \\
&       3 & $  16.8\pm  0.5$ & $  0.54\pm  0.01$ & & & &  \\
&       4 & $   4.1\pm  0.3$ & $  1.13\pm  0.02$ & & & & \\
 Sept.  20 & & & & 0.34  & 1.17$\times$0.33 &  $-26.7^\circ$ & & & 0.14 & 2.8$\times$0.96 & $-29.8^\circ$  \\
&       1 & $  42.8\pm  0.7$ & -  & & & &  $  99.0\pm  1.2$ & - \\
&       2 & $  22.4\pm  0.7$ & $  0.32\pm  0.02$ & & & & $  69.6\pm  1.2$ & $  0.56\pm  0.01$\\
&       3 & $   0.6\pm  0.7$ & $  0.87\pm  0.63$ & & & & \\
&       4 & $   3.7\pm  0.8$ & $  1.27\pm  0.13$ & & & & \\
 Sept.  26 & & & & 0.24  & 0.76$\times$0.29 &  $-16.4^\circ$& & & 0.10 &  2.4$\times$0.75 & $-23.3^\circ$\\
&       1 & $  77.9\pm  0.6$ & - &  & & & $  95.5\pm  1.0$ & -  \\
&       2 & $  37.1\pm  0.7$ & $  0.24\pm  0.01$ &  & & & $  61.4\pm  1.1$ & $  0.48\pm  0.01$ \\
&       3 & $   5.0\pm  0.4$ & $  0.83\pm  0.03$ & & & & \\
&       4 & $   5.2\pm  0.4$ & $  1.34\pm  0.02$ & & & & \\
Oct. 2 & & & & 0.17  & 0.77$\times$0.27 &  $-14.7^\circ$ &  & &  0.16 &  2.3$\times$0.74 & $-13.2^\circ$  \\
&       1 & $ 159.4\pm  0.7$ &  - &  & & & $  94.9\pm  0.4$ & - \\
&       2 & $  31.3\pm  0.6$ & $  0.19\pm  0.01$  & & & & $  65.1\pm  0.4$ & $  0.47\pm  0.01$ \\
&       3 & $   8.0\pm  0.4$ & $  0.46\pm  0.02$ & & & & \\
&       4 & $   2.5\pm  0.3$ & $  1.48\pm  0.04$ & & & & \\
Jan. 16 & & & & 0.32  & 0.43$\times$0.38 &  $36.0^\circ$ & & & 0.22 &  1.1$\times$0.93 & $20.8^\circ$\\
&       1 & $  47.8\pm  0.8$ &  - & & & & $  59.3\pm  0.8$ & - \\
&       2 & $  11.9\pm  0.8$ & $  0.24\pm  0.03$ &  & & & $  17.1\pm  0.8$ & $  0.79\pm  0.02$ \\
&       3 & $   1.7\pm  0.4$ & $  1.24\pm  0.10$ & & & & \\
&       4 & $   1.3\pm  0.4$ & $  1.79\pm  0.13$ & & & & \\

\\

\hline\hline
& & \multicolumn{5}{c}{ 15 GHz } & & & &  \\
\hline 
Jan. 16 & & & & 0.15 & 0.62$\times$0.56 & $-0.6^\circ$ \\
&       1 & $  48.1\pm  0.7$ & -& & & &  \\
&       2 & $  17.3\pm  0.7$ & $  0.29\pm  0.02$& & & &  \\
&       3 & $   2.3\pm  0.4$ & $  1.26\pm  0.07$ & & & & \\
&       4 & $   1.9\pm  0.4$ & $  1.67\pm  0.08$ & & & & \\

\enddata
\tablecomments{ This table shows all the best fit parameters to the fit to the VLBA data.  We assumed point sources for each component, fitting in the UV plane. The FWHM ($\theta_{\rm FWHM}$) and position angles ($\theta_{\rm PA}$) to that of the restoring beam are given for each observation along with the image rms ($\sigma_{\rm rms}$). The 23 GHz and 15 GHz observations were fit with four components, while the 8.4 GHz were fit with two components. All distances are measured relative to the brightest component, i.e., the core, and the component with the largest distance is taken as the ``knot'' in our analysis.}
\end{deluxetable}


\end{document}